\documentclass[preprint,superscriptaddress]{revtex4-2}
\usepackage{qcircuit}

\newcommand{\ket}[1] {\left| #1 \right\rangle}

\begin{document}

\title{Quantum state preparation protocol for encoding classical data into the amplitudes of a quantum information processing register's wave function}

\author{Sahel Ashhab}
\affiliation{Advanced ICT Research Institute, National Institute of Information and Communications Technology (NICT), 4-2-1, Nukui-Kitamachi, Koganei, Tokyo 184-8795, Japan}

\begin{abstract}
We present a protocol for encoding $N$ real numbers stored in $N$ memory registers into the amplitudes of the quantum superposition that describes the state of $\log_2N$ qubits. This task is one of the main steps in quantum machine learning algorithms applied to classical data. The protocol combines partial CNOT gate rotations with probabilistic projection onto the desired state. The number of additional ancilla qubits used during the implementation of the protocol, as well as the number of quantum gates, scale linearly with the number of qubits in the processing register and hence logarithmically with $N$. The average time needed to successfully perform the encoding scales logarithmically with the number of qubits, in addition to being inversely proportional to the acceptable error in the encoded amplitudes. It also depends on the structure of the data set in such a way that the protocol is most efficient for non-sparse data.
\end{abstract}

\maketitle

Quantum computing devices have made great progress towards the construction of a quantum computer whose computing power exceeds that of any existing classical computer \cite{Ladd,Buluta,Kjaergaard}. In particular, a clear quantum advantage over classical computers was recently demonstrated using superconducting devices \cite{Arute,Wu}. Multi-order-of-magnitude increases in the number of qubits and computing power are expected in the coming few years.

On the software side, new algorithms are continually being developed for future quantum computers \cite{Montanaro,Jordan}. In particular, as machine learning techniques become increasingly prevalent, researchers are exploring the potential for quantum computers to offer a computational advantage using similar techniques \cite{Biamonte,SchuldPetruccione}. There have been a large number of proposals for using quantum computers to perform machine learning tasks. There have also been a few proof-of-principle experimental demonstrations of such tasks \cite{Cai,Li,Havlicek,Gianani}.

Quantum machine learning algorithms operate on data stored in the form of a quantum superposition in the state of a quantum information processing register. There are cases where the initial state can be encoded easily into the processing unit for machine learning processing. For example, the data could be a quantum state that results from easily reproducible quantum dynamics, e.g.~a quantum simulation of a physical system. In this case it could be practically impossible to translate this data into classical form (because of the exponentially large Hilbert space) but easy to take the prepared quantum state and perform quantum machine-learning analysis on it. The situation is starkly different when dealing with input data that is provided in classical form and does not necessarily have any relation to quantum mechanical quantities. Assuming that the data is described by a set of $N$ real numbers $\{c_0, c_1, ..., c_{N-1}\}$, one first needs to encode this data into the quantum state of a quantum register. In this case, the step of encoding the classical data into the quantum processor can be the most challenging step in running the machine-learning algorithm.

A conceptually natural encoding of the data, which is used for example in the quantum support vector machine \cite{Rebentrost} and allows a straightforward evaluation of the distance between two data points, is amplitude encoding. This encoding can be described as preparing the state
\begin{equation}
\ket{\Psi} = \frac{1}{\mathcal{N}} \sum_{k=0}^{N-1} c_k \ket{k},
\end{equation}
where $\ket{k}$ is the $n$-qubit state with the integer $k$ expressed in the binary representation, and $n$ is the smallest integer larger than $\log_2N$. For example $\ket{k=5}$ in a three-qubit system would correspond to the state $\ket{101}$. This encoding step in some sense compresses $2^n$ data values such that they are stored in the quantum state of $n$ qubits. The denominator $\mathcal{N}$ is a normalization factor. Here it is given by $\mathcal{N}=\sum_k |c_k|^2$. However, we shall use the same symbol as a generic normalization factor below.

The task of encoding classical data into a quantum register is closely related to the problem of quantum state preparation, which has been studied by several authors in the past two decades \cite{GroverStatePreparation,Grover2002,Kaye,Shende,Mottonen,Soklakov,Wang,Park} and is also closely related to the study of quantum random access memory (qRAM) in more recent literature \cite{Giovannetti,Arunachalam,DiMatteo}. Early studies on this topic showed that one can perform state preparation of a general $n$-qubit state using a sequence of $\sim 2^n$ single- and two-qubit gates \cite{GroverStatePreparation,Kaye,Shende}. Later proposals showed improved, though still exponential, resource scaling \cite{Mottonen}. These results are consistent with intuition based on the fact that a general $n$-qubit state is defined by $2^n$ complex basis state amplitudes, with one normalization constraint and one irrelevant overall phase. The state-preparation gate sequence must therefore contain at least $2^{(n+1)}-2$ adjustable parameters to be able to access any point in the $n$-qubit Hilbert space, which leads to the conclusion that $\sim 2^n$ single- and two-qubit gates are needed to perform state preparation with an arbitrary target state. Other proposals showed that polynomial scaling in $n$ can be achieved for special cases depending on the structure of the data, e.g.~if the number of basis states in the superposition is small \cite{Park} or if the superposition contains only basis states with a certain number of zeros and ones \cite{Wang,DiMatteo}. Other proposals demonstrated polynomial scaling in $n$ based on the assumption that a certain oracle that contains the data as controllable parameters can be implemented efficiently \cite{Grover2002,Soklakov}. A protocol integrating qRAM and state preparation steps was described in Ref.~\cite{SchuldPetruccione}. However, this protocol also does not specify the details of the oracle implementation. Another recent direction of research in this area is approximate encoding, where adaptive learning techniques have been proposed to optimize the performance of encoding protocols for a fixed amount of resources \cite{Schuld,Lloyd,Zoufal,Nakaji}.

Here we present a protocol that performs amplitude encoding efficiently when an exponentially large number of amplitudes are present in the data to be encoded, explicitly describing the quantum operations used in the protocol. The protocol is based on the use of partial CNOT gates and probabilistic measurement-induced projection. The steps used in this protocol are similar to those used in Grover's state preparation protocol \cite{GroverStatePreparation,Sanders}. As we shall see below, the protocol is most efficient when a large fraction of the values in the data set are of the same overall scale, while the protocol becomes increasingly inefficient as we approach the sparse limit where most of the data are zeros or negligibly small relative to the largest value in the data set.

As our starting point, we assume that the numbers $c_k$ are stored in $N$ memory registers, each as an integer with a length of $L$ bits, where $L$ sets the accuracy of the numbers $c_k$. Each memory register therefore contains $L$ bits that we call the value bits. We also assume that each memory register contains $n$ additional bits that encode the integer $k$. We refer to this part of the memory register as the index bits. The state of each memory register can therefore be expressed as $\ket{k,c_k}$, with a total of $n+L$ bits in each register. For example, if we take a data set that contains four elements, each of which contains five binary digits, the entire memory is described by the state
\begin{eqnarray}
& & \ket{0, 0, c_{0,0}, c_{0,1}, c_{0,2}, c_{0,3}, c_{0,4}} \otimes \nonumber \\ & & \ket{0, 1, c_{1,0}, c_{1,1}, c_{1,2}, c_{1,3}, c_{1,4}} \otimes \nonumber \\ & & \ket{1, 0, c_{2,0}, c_{2,1}, c_{2,2}, c_{2,3}, c_{2,4}} \otimes \nonumber \\ & & \ket{1, 1, c_{3,0}, c_{3,1}, c_{3,2}, c_{3,3}, c_{3,4}},
\end{eqnarray}
where $c_{k,l}$ (with $l=0,1,2,...,L-1$) are the individual bits in the bit string that encodes the value $c_k$. In practice, it might not be necessary to encode the index bits in a dedicated part of the memory, as the hardware might be designed to correctly identify the index of any value stored in the memory based on the location where the value is stored. It should be noted that the memory registers are in a classical state that does not involve any quantum superposition, and they remain in the same state throughout the protocol. However, they must be physically isolated from the environment to prevent the environment from knowing which memory register is accessed. The reason is that amplitude encoding relies on having a quantum superposition over all values of $k$ and each memory register is accessed in one branch of the superposition. If the information about which memory register is accessed leaks to the environment, this information would show that only one memory register (with a specific value of $k$) was accessed, and the state of the quantum computer would collapse to a state with a single value of $k$. For this reason we shall refer to the physical components that hold the memory bits as qubits, even though they remain in a classical state throughout the protocol.

To provide insight into the basic idea of the protocol, we start by presenting its first steps in the simple case where the data set contains only two numbers, $c_0$ and $c_1$, such that these two numbers are to be encoded into a single-qubit processing register, to which we shall refer as the CPU register. The memory is therefore in the state $\ket{0,c_0} \otimes \ket{1,c_1}$. We initialize the CPU qubit in the state $(\ket{0}+\ket{1})/\sqrt{2}$. In addition to the CPU register, we introduce an ancilla qubit, the flag qubit, initialized in the state $\ket{0}$. Hence the combined system is initially in the state
\begin{equation}
\frac{1}{\sqrt{2}} \left( \ket{0} + \ket{1} \right) \otimes \ket{0} \otimes \prod_{k=0}^{1} \ket{k,c_k},
\end{equation}
where the first ket is that of the CPU qubit, the second ket is that of the flag qubit, and the product of kets on the right describes the state of the two memory registers. 

With this initial state, we seek an operation that will modify the amplitudes in the quantum superposition and make them proportional to the corresponding numbers $c_k$. Considering only one of the two computational basis states and its corresponding memory register, an operation that produces an amplitude that is approximately proportional to $c_k$ is a rotation by an angle that is proportional to the value in the memory register. If such a rotation is applied to the flag qubit, it causes the transformation:
\begin{equation}
\ket{k} \otimes \ket{0} \otimes \ket{k,c_k} \rightarrow \ket{k} \otimes \left( \cos\frac{c_k}{\mathcal{R}} \ket{0} + \sin\frac{c_k}{\mathcal{R}} \ket{1} \right) \otimes \ket{k,c_k}.
\label{Eq:SingleFiniteCoefficientTransformation}
\end{equation}
The parameter $\mathcal{R}$ is a constant that the user can choose freely, and it should be chosen to be much larger than $|c_k|$, such that $\sin(c_k/\mathcal{R})$ is approximately equal to $c_k/\mathcal{R}$, as we shall discuss in more detail below.

The above-described rotation, which needs to be implemented for one value of $k$, can be described by the unitary operation $U_k=e^{-i\sigma_{y}^{F} c_k/\mathcal{R}}$, where $\sigma_{y}^{F}$ is a Pauli operator that flips the states $\ket{0}$ and $\ket{1}$ of the flag qubit. The operator $U_k$ can be expressed more explicitly in terms of the individual bits $c_{k,l}$ that comprise the number $c_k$:
\begin{equation}
U_k = \exp \left\{ -i \sigma_{y}^{F} \left( \sum_{l=0}^{L-1} \frac{c_{k,l} 2^{L-l-1}}{\mathcal{R}} \right) \right\}.
\end{equation}
Note that, depending on the convention used for the scale of the numbers $c_k$, there could be an overall scale factor, e.g.~$2^L$, inside the exponent that we have omitted. Any such factor can be absorbed into the constant $\mathcal{R}$ and hence does not affect the protocol. In an efficient encoding protocol, the externally applied controls should not explicitly depend on the data values. Instead, these values should be retrieved from the memory and used to implement the necessary rotations on the flag qubit via a unitary operator that does not explicitly contain the numbers $c_k$. To demonstrate how this goal can be achieved, we first turn the numbers $c_{k,l}$ in $U_k$ into operators. We therefore rewrite the operator $U_k$ in the form
\begin{equation}
U_k = \prod_{l=0}^{L-1} \exp \left\{ -i \frac{2^{L-l-2}}{\mathcal{R}} \sigma_{y}^{F} \otimes \left( 1 - \sigma_z^{{\rm Memory},k,l} \right) \right\},
\label{Eq:U_k_product}
\end{equation}
where $\sigma_z^{{\rm Memory},k,l}$ is the $z$ Pauli operator value for the $l$th qubit in the value part of the $k$th memory register, i.e.~$\sigma_z^{{\rm Memory},k,l}=1$ if $c_{k,l}=0$ and $\sigma_z^{{\rm Memory},k,l}=-1$ if $c_{k,l}=1$. We have therefore turned each term in $U_k$ into a partial CNOT operation in which one of the memory qubits is the control qubit and the flag qubit is the target qubit. The operation $U_k$ is composed of $L$ such operations, one for each value of $l$. Each one of these partial CNOT operations can be implemented by using Rabi oscillation dynamics in the flag qubit conditioned on the state of one bit in the memory register. Such conditional oscillation dynamics is in fact common in physical implementations of quantum information devices. The CNOT gate is usually generated by creating conditions for a control-qubit-dependent resonance, which results in Rabi oscillations in the target qubit conditioned on the state of the control qubit. By setting the pulse time such that half a Rabi oscillation is completed, a CNOT gate is implemented. By varying the pulse time, a partial CNOT gate can be implemented. Specifically, by setting the pulse time to obtain a rotation angle of $2^{L-l}/\mathcal{R}$ and using the $l$th memory qubit as the control qubit, the rotation corresponding to the $l$th term in Eq.~(\ref{Eq:U_k_product}) can be realized. An example of such an implementation of the CNOT gate dynamics that is one of the standard two-qubit gate protocols for superconducting qubits is described in Ref.~\cite{DeGroot}. Each one of the rotations in the product in Eq.~(\ref{Eq:U_k_product}) is controlled by  one of the memory qubits. Since all of these operations commute with each other, they can all be implemented simultaneously.

The role that the index bits can play in the protocol becomes clear when we consider how to simultaneously implement the two transformations corresponding to the two values of $k$. In the absence of another mechanism to identify which memory value bits to use in the implementation of $U_k$, a condition can be incorporated into $U_k$ in the form of the operator $(1+\sigma_z^{\rm CPU}\otimes\sigma_z^{{\rm Memory \ index}, k})/2$, where $\sigma_z^{\rm CPU}$ and $\sigma_z^{{\rm Memory \ index}, k}$ are defined similarly to $\sigma_z^{{\rm Memory},k,l}$ but for the CPU qubit and the $k$th memory index qubit, respectively. This combined operator is equal to 1 if the value of the memory index qubit matches the value of the CPU qubit in the computational basis, i.e.~if they are in the state $\ket{0}\otimes\ket{0}$ or the state $\ket{1}\otimes\ket{1}$, and the operator is equal to 0 otherwise. We note here that an operation on a target qubit that is conditioned on the matching between two control qubits can be performed by first mapping the parity $P$ of the two control qubits onto an ancilla qubit that is then used as the control qubit in a controlled operation \cite{Moore}, as illustrated in the following diagram: 

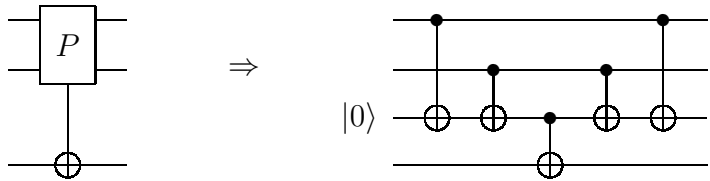
\centerline{
\Qcircuit @C=1em @R=.7em {
& \multigate{1}{P} & \qw & & & \ctrl{2} & \qw & \qw & \qw & \ctrl{2} & \qw \\
& \ghost{P} & \qw & \push{\rule{2.0em}{0em}\Rightarrow\rule{3.0em}{0em}} & & \qw & \ctrl{1} & \qw & \ctrl{1} & \qw & \qw \\
& \qwx & & & \lstick{\ket{0}} & \targ & \targ & \ctrl{1} & \targ & \targ & \qw \\
& \targ \qwx & \qw & & & \qw & \qw & \targ & \qw & \qw & \qw
}
}

\

\noindent With the matching condition incorporated into $U_k$, even if the combined system is controlled externally in such a way that all memory registers (corresponding to the different values of $k$) are accessed simultaneously on an equal footing, this matching condition ensures that the appropriate values $c_{k,l}$ are used in the $k$th branch of the quantum superposition when implementing the unitary operation. The operator $U_k$ can then be replaced by the operator
\begin{equation}
U = \prod_{k=0}^{1} \prod_{l=0}^{L-1} \exp \left\{ -i \frac{2^{L-l-3}}{\mathcal{R}} \sigma_{y}^{F} \otimes \left( 1 - \sigma_z^{{\rm Memory},k,l} \right) \otimes \left( 1 + \sigma_z^{\rm CPU} \otimes \sigma_z^{{\rm Memory \ index},k} \right) \right\},
\end{equation}
which performs the necessary transformations for both values of $k$. It should be emphasized here that the product over $k$ and the additional conditioning operator inside the exponential do not add serious complications to the implementation of $U$, e.g.~in terms of resource scaling or the need to perform a separate operation for each value of $k$. As explained above, the product over $k$ simply means that all the memory registers, which are separate physical objects, are accessed in the process. Physically only a single query operation, which has the ability to access any part of the memory, is applied. The part of the memory that responds to the query is determined by the state of the CPU qubit in the computational basis. As for the additional operator inside the exponential, although the operator $U$ is now a multi-qubit operator in which the flag qubit rotation is conditioned on the memory value qubits as well as the matching of the CPU and memory index qubits (both of which become qubit strings when $n>1$), the different condition terms can be efficiently mapped onto a single qubit that is used as the only control qubit in the implementation of the controlled rotation \cite{Nielsen,Selinger}. More specifically, if we are given $K$ control qubits and we wish to implement an operation on a target qubit conditioned on all the control qubits being in the state $\ket{1}$, we can use $K/2$ ancilla qubits initialized in the state $\ket{0}$ and perform $K/2$ Toffoli gates to obtain $K/2$ control qubits instead of the original $K$ control qubits. Next we use $K/4$ additional qubits and repeat the process to halve the number of control qubits once again. We repeat this process $\log_2 K$ times, using a total of $K-1$ ancilla qubits, to obtain a single control qubit that is in the state $\ket{1}$ if and only if all the original control qubits are in the state $\ket{1}$. After the controlled operation on the target qubit (which here is the flag qubit), the ancilla qubits can be returned to their initial state (i.e.~$\ket{0}$ for all the ancilla qubits) by reversing the above process, such that the ancilla qubits are disentangled from the rest of the system, which is needed to prepare the desired state in the CPU qubit \cite{JWfootnote}. As an example, the procedure for implementing a multi-qubit-controlled operation for four control qubits is illustrated in the following diagram:

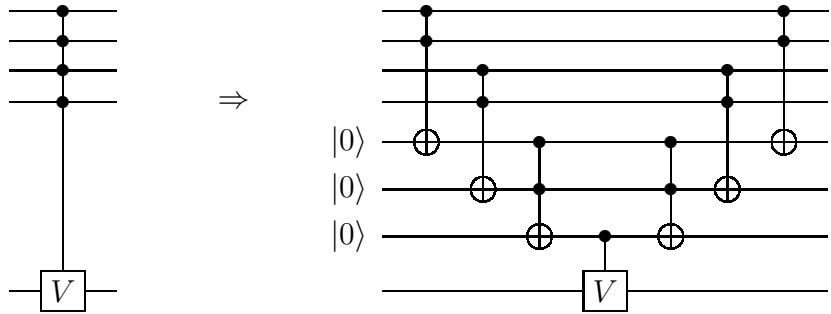
\centerline{
\Qcircuit @C=1em @R=.7em {
& \ctrl{7} & \qw & & & \ctrl{4} & \qw & \qw & \qw & \qw & \qw & \ctrl{4} & \qw \\
& \ctrl{6} & \qw & & & \ctrl{3} & \qw & \qw & \qw & \qw & \qw & \ctrl{3} & \qw \\
& \ctrl{5} & \qw & & & \qw & \ctrl{3} & \qw & \qw & \qw & \ctrl{3} & \qw & \qw \\
& \ctrl{4} & \qw & \push{\rule{2.0em}{0em}\Rightarrow\rule{3.0em}{0em}} & & \qw & \ctrl{2} & \qw & \qw & \qw & \ctrl{2} & \qw & \qw \\
& & & & \lstick{\ket{0}} & \targ & \qw & \ctrl{2} & \qw & \ctrl{2} & \qw & \targ & \qw \\
& & & & \lstick{\ket{0}} & \qw & \targ & \ctrl{1} & \qw & \ctrl{1} & \targ & \qw & \qw \\
& & & & \lstick{\ket{0}} & \qw & \qw & \targ & \ctrl{1} & \targ & \qw & \qw & \qw \\
& \gate{V} & \qw & & & \qw & \qw & \qw & \gate{V} & \qw & \qw & \qw & \qw
}
}

\

\noindent We also note that an alternative implementation of multi-qubit-controlled operations, relying on the use of qutrits instead of qubits and not requiring any ancillae, was proposed recently in Ref.~\cite{Inada}.

Application of the operation $U$ leads to the following transformation in the quantum state of the system
\begin{eqnarray}
\frac{1}{\sqrt{2}} \left( \ket{0} + \ket{1} \right) \otimes \ket{0} \otimes \prod_{k=0}^{1} \ket{k,c_k} & \rightarrow & \frac{1}{\sqrt{2}} \Bigg\{
\ket{0} \otimes \left( \cos\frac{c_0}{\mathcal{R}} \ket{0} + \sin\frac{c_0}{\mathcal{R}} \ket{1} \right) + 
\nonumber \\ & & \hspace{1cm}
\ket{1} \otimes \left( \cos\frac{c_1}{\mathcal{R}} \ket{0} + \sin\frac{c_1}{\mathcal{R}} \ket{1} \right)
\Bigg\} \otimes \prod_{k=0}^{1} \ket{k,c_k}.
\label{Eq:SingleQubitTransformationFirstStep}
\end{eqnarray}

It is now convenient to turn to the case of a multi-qubit CPU register. The CPU register is initialized in the state $2^{-n/2}(\ket{0}+\ket{1})^{\otimes n}$, which can alternatively be expressed as $2^{-n/2}\sum_{k=0}^{2^n-1} \ket{k}$. Similarly to the single-qubit case, the protocol proceeds by implementing a conditional rotation on the flag qubit controlled by the CPU and memory registers:
\begin{eqnarray}
\left( \sum_{k=0}^{2^n-1} \frac{\ket{k}}{2^{n/2}} \right) \otimes \ket{0} \otimes \prod_{k=0}^{N-1} \ket{k,c_k} & \rightarrow & \frac{1}{2^{n/2}}\Bigg\{
\ket{0} \otimes \left( \cos\frac{c_0}{\mathcal{R}} \ket{0} + \sin\frac{c_0}{\mathcal{R}} \ket{1} \right) +
\nonumber \\ & & \hspace{1.2cm}
\ket{1} \otimes \left( \cos\frac{c_1}{\mathcal{R}} \ket{0} + \sin\frac{c_1}{\mathcal{R}} \ket{1} \right) +
\nonumber \\ & & \hspace{1cm}
\hspace{3cm} \vdots
\nonumber \\ & & \hspace{1.2cm}
\ket{2^n-1} \otimes \left( \cos\frac{c_{2^n-1}}{\mathcal{R}} \ket{0} + \sin\frac{c_{2^n-1}}{\mathcal{R}} \ket{1} \right)
\Bigg\}
\nonumber \\ & & \hspace{0cm}
\otimes \prod_{k=0}^{N-1} \ket{k,c_k}.
\label{Eq:MultiQubitTransformationFirstStep}
\end{eqnarray}
This transformation is implemented using the operator
\begin{equation}
U = \prod_{k=0}^{N-1} \prod_{l=0}^{L-1} \exp \left\{ -i \frac{2^{L-l-3}}{\mathcal{R}} \sigma_{y}^{F} \otimes \left( 1 - \sigma_z^{{\rm Memory},k,l} \right) \otimes \prod_{m=1}^{n} \left( 1 + \sigma_z^{{\rm CPU},m} \otimes \sigma_z^{{\rm Memory \ index},k,m} \right) \right\},
\label{Eq:MultiQubitU}
\end{equation}
where the index $m$ labels the $n$ qubits in the CPU register and in each memory index register. The product over $m$ is equal to 1 if the state of the CPU register matches that of the memory index register and is equal to 0 otherwise. As a result, the operator $U$ uses the appropriate values, i.e.~$c_{k,l}$, in the $k$th branch of the quantum superposition of computational basis states.

The implementation of the operator $U$ in Eq.~(\ref{Eq:MultiQubitU}) proceeds as follows: first, the information about the matching between the CPU qubits and memory index qubits is mapped onto $n$ ancilla qubits, which we call the parity qubits. Then $n-1$ additional ancilla qubits are used to compress the information in the $n$ parity qubits into a single qubit. Then this single qubit is used as a control qubit to implement the rotations in the flag qubits based on the values of the memory value bits. There are $L$ of these rotations, one for each value of $l$. All of these operations can be performed simultaneously. It should also be noted that each one of these rotations is a Toffoli gate with two control qubits: one memory value qubit and the one ancilla qubit that encodes the full CPU-memory-index matching condition. After the controlled rotations, the ancilla qubits are returned to their initial state.

After performing the above transformation (described by the operator $U$), we perform a measurement on the flag qubit. If it is found to be in the state $\ket{1}$, the system is projected onto the state
\begin{eqnarray}
\frac{1}{\mathcal{N}} \Bigg\{
\sin\frac{c_0}{\mathcal{R}} \ket{0} +
\sin\frac{c_1}{\mathcal{R}} \ket{1} +
\cdots
\sin\frac{c_{2^n-1}}{\mathcal{R}} \ket{2^n-1}
\Bigg\} \otimes \ket{1} \otimes \prod_{k=0}^{N-1} \ket{k,c_k}.
\nonumber
\label{Eq:MultiQubitTransformationAfterMeasurementStep}
\end{eqnarray}
The state of the CPU register is now disentangled from that of the flag qubit. If we set $\mathcal{R}$ to a sufficiently large value such that $|c_k/\mathcal{R}|\ll 1$ for all values of $k$, the state of the CPU register can be expressed as
\begin{eqnarray}
\ket{\Psi_{\rm Final}^{\rm CPU}} = \frac{1}{\mathcal{N}} \Bigg\{
\frac{c_0}{\mathcal{R}} \ket{0} +
\frac{c_1}{\mathcal{R}} \ket{1} +
\cdots +
\frac{c_{2^n-1}}{\mathcal{R}} \ket{2^n-1}
\Bigg\} + \sum_k \mathcal{O}\left( \frac{c_k}{\mathcal{R}} \right)^3,
\label{Eq:MultiQubitTransformationAfterFinalApproximation}
\end{eqnarray}
which, to lowest order, is the desired state. We shall comment on the deviation term shortly.

We now consider the resources required for the implementation of the protocol. In the steps described above, we introduced $2n$ extra qubits in addition to the CPU and memory registers. These extra qubits are: one flag qubit, $n$ ancilla qubits to temporarily store the parity information, and $n-1$ additional ancilla qubits for the implementation of the $n$-qubit-controlled rotations, i.e.~for compressing the parity information into a single qubit. The number of single- and two-qubit gates performed during the protocol is the sum of two terms: $\sim n$ gates needed to prepare the single control qubit and $\sim L$ gates needed to perform the controlled rotations on the flag qubit. The total number of gates therefore scales linearly with the larger of the two parameters $n$ and $L$. As explained above, the $L$ operations involving different values of $l$ can be performed simultaneously. The $n$ operations needed to prepare the parity qubits can also be performed simultaneously. The step of preparing a single control qubit from $n$ control qubits can be partially parallelized. Specifically, the depth, i.e.~the minimum number of steps when as many single- and two-qubit gates as possible are performed simultaneously, scales only as $\log_2 n$, since each step in the control qubit compression procedure halves the number of control qubits. In addition to these scaling laws, the resource requirements will depend on the success probability of the measurement step, i.e.~the probability that the flag qubit will be found in the state $\ket{1}$. The time needed to successfully prepare the desired state will, on average, be proportional to the inverse of the success probability. The probability that the flag qubit will be in the state $\ket{1}$ just before the measurement is given by
\begin{equation}
P_{\rm Success} = \frac{1}{2^n} \sum_k \left| \sin\frac{c_k}{\mathcal{R}} \right|^2.
\label{Eq:SuccessProbability}
\end{equation}
The success probability therefore depends on $c_k$, which are the data values that we wish to encode, as well as the parameter $\mathcal{R}$, which is a variable that we can set freely.

First we consider the parameter $\mathcal{R}$. If we want to maximize $P_{\rm Success}$ while ignoring all other considerations, we choose a small value of $\mathcal{R}$, ideally a value comparable to $c_k$. However, $\mathcal{R}$ must be much larger than the largest value of $|c_k|$, to which we refer as $c_{\rm max}$, to make sure that the approximation in Eq.~(\ref{Eq:MultiQubitTransformationAfterFinalApproximation}) is valid for all values of $k$. The question then is how small we can take $\mathcal{R}$ before we start having a nonnegligible deviation from the ideal final state. The coefficient of the state $\ket{k}$ in the final state of the protocol is $\sin(c_k/\mathcal{R})$ instead of being $c_k/\mathcal{R}$. The relative error is therefore given by
\begin{equation}
\epsilon_k = \frac{\sin(c_k/\mathcal{R})}{c_k/\mathcal{R}} - 1 = - \frac{1}{6} \left( \frac{c_k}{\mathcal{R}} \right)^2 + \mathcal{O}(\frac{c_k}{\mathcal{R}})^4.
\end{equation}
It is worth noting here that this expression for the error is a conservative estimate: because all values of $\epsilon_k$ are negative, the renormalization factor in Eq.~(\ref{Eq:MultiQubitTransformationAfterFinalApproximation}) will at least partially suppress the difference between the coefficients in the prepared state and in the desired state. If we set a maximum acceptable relative error of $\epsilon$ in any individual value in the data, $\mathcal{R}$ must be chosen such that $c_{\rm max}/\mathcal{R} \leq \sqrt{6\epsilon}$. In other words $\mathcal{R}\geq c_{\rm max}/\sqrt{6\epsilon}$. Substituting this inequality into Eq.~(\ref{Eq:SuccessProbability}), we obtain the inequality
\begin{eqnarray}
P_{\rm Success} & \leq & \frac{1}{2^n} \sum_k \left| \sin\frac{c_k\sqrt{6\epsilon}}{c_{\rm max}} \right|^2
\nonumber \\
& \leq & \frac{6\epsilon}{2^n} \sum_k \left| \frac{c_k}{c_{\rm max}} \right|^2.
\label{Eq:SuccessProbabilityInequality}
\end{eqnarray}
If we assume that the optimal value of $\mathcal{R}$ is chosen, i.e.~if $\mathcal{R}$ is much larger than $c_{\rm max}$ but does not diverge with increasing $n$ such that it affects the scaling law, $P_{\rm Success}$ will be on the order of the right-hand side of Eq.~(\ref{Eq:SuccessProbabilityInequality}).

To analyze the role that the structure of the data plays in the efficiency of the protocol, it is instructive to define the density (or, in other words, non-sparsity) measure
\begin{equation}
\rho = \frac{1}{2^n} \sum_k \left| \frac{c_k}{c_{\rm max}} \right|^2,
\end{equation}
which allows us to express the optimized success probability as
\begin{equation}
P_{\rm Success} \propto \rho \epsilon.
\end{equation}
The parameter $\rho$ takes its maximum value ($\rho=1$) when all $|c_k|$ are equal, while $\rho \ll 1$ for sparse data, where only a small number of $c_k$ values are on the same order as $c_{\rm max}$. If a substantial fraction of all the numbers $|c_k|$ in the data set are comparable to each other, $\rho$ will be on the order of one. This situation is desirable for maximizing $P_{\rm Success}$. If on the other hand only a small fraction of the data set is nonnegligible in comparison with the largest value $c_{\rm max}$, $\rho$ will be much smaller than one, and $P_{\rm Success}$ will be especially small. In the worst case scenario, when the number of large $|c_k|$ values does not grow with the size of the data set, $\rho$ decreases exponentially with $n$. The latter case corresponds to a very sparse data set. The protocol is therefore well suited for dense data sets where a significant fraction of the data is on the same order as the maximum value. The more sparse the data, the less efficient the protocol. Note that in order to make the assessment about the value of $\rho$ we need to have some basic information about the overall properties of the data set. Note also that some of this information can be measured. For example, $P_{\rm Success}$ can be measured and used to obtain an estimate for $\sum_k |c_k|^2$.

Combining the success probability ($\propto \rho\epsilon$) with the time needed to implement the steps of the protocol ($\propto \log n$), we obtain the scaling law for the time needed to successfully prepare the encoded state:
\begin{equation}
T \propto \frac{\log n}{\rho\epsilon}.
\end{equation}
Importantly, provided that the data is not sparse (i.e.~the parameter $\rho$ is not exponentially small) as explained above, there are no exponential factors in the resource scaling laws. The protocol can therefore be considered efficient in this case. It is worth noting here that dense data sets with an exponentially large number of elements are generally considered the most difficult ones for classical algorithms, because these algorithms operate on each element separately, leading to an exponential scaling in the required resources.

A few comments are in order at this point. The numbers $c_k$ are usually treated as real numbers, while the coefficients in a quantum superpositions can in general be complex. As a result, one could compress the data further by taking advantage of the complex-number nature of the quantum superposition. However, we ignore this possibility in this work. The adjustment of the protocol to take advantage of this fact should be relatively straightforward. However, the factor-of-two resource savings translates into a reduction of the required qubit number by one qubit, which is in some sense minimal and not worth complicating the physical picture.

By setting a sufficiently large value for $\mathcal{R}$ and implementing the protocol a large number of times, the success probability (heralded by the flag qubit being found in the state $\ket{1}$) gives a good approximation for the sum $\sum_k \left| c_k \right|^2$. However, it is not sufficient to rely on $P_{\rm Success}$ to estimate the scale of $c_{\rm max}$ and set $\mathcal{R}$. In particular, if one or a few $|c_k|$ values are very large compared to the bulk of the $|c_k|$ values, and the latter are all at the same scale, the behavior of $P_{\rm Success}$ as a function of $\mathcal{R}$ would be governed by the scale of the majority of $|c_k|$ values. One might then unknowingly set $\mathcal{R}$ to a value that is smaller than $c_{\rm max}$. The controlled rotation dynamics for the states $\ket{k}$ with large $|c_k|$ values will then take the form of many Rabi oscillations and the final state after the controlled rotation will leave the state $\ket{k}$ with a probability somewhere between 0 and 1, as opposed to $|c_k/\mathcal{R}|^2$ which is much larger than 1 under this scenario. On the other hand, if the few large values of $|c_k|$ are treated as outliers that should not dominate the final results of the computation, this approach naturally suppresses the values of the outliers and, as a result, their contribution to the final output.

Another point concerns how to do the resource counting. It might seem at first sight that there is exponential scaling in resources because the protocol involves $2^n$ memory registers, i.e.~a number that scales exponentially with $n$. However, this does not imply that the protocol is not efficient. The $\sim 2^n$ memory qubits store the input data. Therefore their large number is simply a reflection of the fact that we are given an extremely large amount of input data for the computation. The more appropriate comparison is to say that, given this amount of data, a classical algorithm requires resources that scale as $2^n$ (i.e.~$N$), while the quantum protocol could require resources that scale only as $n$ (i.e.~$\log_2N$), hence an exponential speedup. A related point is that in implementing the conditional operations (Eq.~\ref{Eq:U_k_product}) an exponentially large number of operations are implemented. However, considering that the hardware is normally set up such that any memory register can be accessed, the $N$ different memory registers are accessed simultaneously in different branches of the quantum superposition with the relevant memory register activated by the state of the CPU register. This situation is somewhat similar to that encountered in Grover's database search algorithm \cite{GroverSearch} or qRAM protocols \cite{DiMatteo}, where all the data stored in the memory are queried simultaneously. One can therefore say that we are assuming a setup that supports such operations.

In this work we focused on the problem of encoding the entire data set, with all different values of $k$, into the CPU register. In some cases, one might be interested in analyzing only a subset of the data, e.g.~the elements for which the index $k$ satisfies a certain condition. It should be possible to incorporate such a condition on $k$ when implementing the conditional operation $U$. The quantum circuit needed to implement such a condition will depend on the nature of the condition defining the subset of interest. A related point is that we assumed a uniform quantum superposition in the initial state of the CPU register, as well as an ancilla flag qubit initialized in the state $\ket{0}$. Considering more general initial states can lead to a richer, and possibly computationally advantageous, variety of final states. We do not consider these possible extensions of our protocol here.

In conclusion, using steps similar to those used in Grover's state preparation algorithm and qRAM protocols, we have developed a protocol for amplitude encoding of classical data into a quantum processing register for quantum machine learning. Provided that the data is not sparse, the protocol is efficient. This proposal addresses one of the main bottlenecks for quantum machine learning algorithms and can be integrated into such algorithms in future quantum computing applications.

We would like to thank Jae Park, Kouichi Semba, Hefeng Wang and Naoki Yamamoto for useful discussions. This work was supported by MEXT Quantum Leap Flagship Program Grant Number JPMXS0120319794.


\begin{thebibliography}{99}

\bibitem{Ladd} T. D. Ladd, F. Jelezko, R. Laflamme, Y. Nakamura, C. Monroe, and J. L. O'Brien, Quantum computers, Nature {\bf 464}, 45 (2010).

\bibitem{Buluta} I. Buluta, S. Ashhab, and F. Nori, Natural and artificial atoms for quantum computation, Rep. Prog. Phys. {\bf 74}, 104401 (2011).

\bibitem{Kjaergaard} M. Kjaergaard, M. E. Schwartz, J. Braum\"uller, P. Krantz, J. I.-J. Wang, S. Gustavsson, and W. D. Oliver, Superconducting qubits: current state of play, Annu. Rev. Condens. Matter Phys. {\bf 11}, 369 (2020).

\bibitem{Arute} F. Arute {\it et al.}, Quantum supremacy using a programmable superconducting, Nature {\bf 574}, 505 (2019).

\bibitem{Wu} Y. Wu {\it et al.}, Strong quantum computational advantage using a superconducting quantum processor, Phys. Rev. Lett. {\bf 127}, 180501 (2021).

\bibitem{Montanaro} A. Montanaro, Quantum algorithms: an overview, npj Quantum Inf. {\bf 2}, 15023 (2016).

\bibitem{Jordan} S. Jordan, Quantum Algorithm Zoo, www.quantumalgorithmzoo.org.

\bibitem{Biamonte} J. Biamonte, P. Wittek, N. Pancotti, P. Rebentrost, N. Wiebe, and S. Lloyd, Quantum machine learning, Nature {\bf 549}, 195 (2017).

\bibitem{SchuldPetruccione} M. Schuld and F. Petruccione, {\it Supervised Learning with Quantum Computers} (Springer, Cham, 2018).

\bibitem{Cai} X.-D. Cai, D. Wu, Z.-E. Su, M.-C. Chen, X.-L. Wang, L. Li, N.-L. Liu, C.-Y. Lu, J.-W. Pan, Entanglement-based machine learning on a quantum computer, Phys. Rev. Lett. {\bf 114}, 110504 (2015).

\bibitem{Li} Z. Li, X. Liu, N. Xu, and J. Du, Experimental realization of a quantum support vector machine, Phys. Rev. Lett. {\bf 114},140504 (2015).

\bibitem{Havlicek} V. Havl\'i\v{c}ek, Antonio D. C\'orcoles, K. Temme, A. W. Harrow, A. Kandala, J. M. Chow, and J. M. Gambetta, Supervised learning with quantum-enhanced feature spaces, Nature {\bf 567}, 209 (2019).

\bibitem{Gianani} I. Gianani, I. Mastroserio, L. Buffoni, N. Bruno, L. Donati, V. Cimini, M. Barbieri, F. S. Cataliotti, and F. Caruso, Experimental quantum embedding for machine learning, arXiv:2106.13835.

\bibitem{Rebentrost} P. Rebentrost, M. Mohseni, and S. Lloyd, Quantum support vector machine for big data classification, Phys. Rev. Lett. {\bf 113}, 130503 (2014).

\bibitem{GroverStatePreparation} L. K. Grover, Synthesis of quantum superpositions by quantum computation, Phys. Rev. Lett. {\bf 85}, 1334 (2000).

\bibitem{Grover2002} L. Grover and T. Rudolph, Creating superpositions that correspond to efficiently integrable probability distributions, arXiv:quant-ph/0208112.

\bibitem{Kaye} P. Kaye and M. Mosca, Quantum networks for generating arbitrary quantum states, Proceedings of the International Conference on Quantum Information (Rochester, New York, 2001).

\bibitem{Shende} V. V. Shende and I. L. Markov, Quantum circuits for incompletely specified two-qubit operators, Quantum Inf. Comput. {\bf 5}, 49 (2005).

\bibitem{Mottonen} M. M\"ott\"onen, J. J. Vartiainen, V. Bergholm, and M. M. Salomaa, Transformation of quantum states using uniformly controlled rotations, Quantum Info. Comput. {\bf 5}, 467 (2005).

\bibitem{Soklakov} A. N. Soklakov and R. Schack, Efficient state preparation for a register of quantum bits, Phys. Rev. A {\bf 73}, 012307 (2006).

\bibitem{Wang} H. Wang, S. Ashhab, and F. Nori, Efficient quantum algorithm for preparing molecular-system-like states on a quantum computer, Phys. Rev. A {\bf 79}, 042335 (2009).

\bibitem{Park} D. K. Park, F. Petruccione, and J.-K. K. Rhee, Circuit-based quantum random access memory for classical data, Sci. Rep. {\bf 9}, 3949 (2019).

\bibitem{Giovannetti} V. Giovannetti, S. Lloyd and L. Maccone, Quantum random access memory, Phys. Rev. Lett. {\bf 100}, 160501 (2008).

\bibitem{Arunachalam} S. Arunachalam, V. Gheorghiu, T. Jochym-O’Connor, M. Mosca and P. V. Srinivasan, On the robustness of bucket brigade quantum RAM, New J. Phys. {\bf 17}, 123010 (2015).

\bibitem{DiMatteo} O. Di Matteo, V. Gheorghiu, and M. Mosca, Fault-tolerant resource estimation of quantum random-access memories, IEEE Trans. Quantum Eng. {\bf 1}, 1 (2020).

\bibitem{Schuld} M. Schuld and N. Killoran, Quantum machine learning in feature Hilbert spaces, Phys. Rev. Lett. {\bf 122}, 040504 (2019).

\bibitem{Lloyd} S. Lloyd, M. Schuld, A. Ijaz, J. Izaac, and N. Killoran, Quantum embeddings for machine learning, arXiv:2001.03622.

\bibitem{Zoufal} C. Zoufal, A. Lucchi and S. Woerner, Quantum generative adversarial networks for learning and loading random distributions, npj Quantum Inf. {\bf 5}, 103 (2019).

\bibitem{Nakaji} K. Nakaji, S. Uno, Y. Suzuki, R. Raymond, T. Onodera, T. Tanaka, H. Tezuka, N. Mitsuda, N. Yamamoto, Approximate amplitude encoding in shallow parameterized quantum circuits and its application to financial market indicator, arXiv:2103.13211.

\bibitem{Sanders} Y. R. Sanders, G. H. Low, A. Scherer, and D. W. Berry, Black-box quantum state preparation without arithmetic, Phys. Rev. Lett. {\bf 122}, 020502 (2019).

\bibitem{DeGroot} P. C. de Groot, J. Lisenfeld, R. N. Schouten, S. Ashhab, A. Lupascu, C. J. P. M. Harmans, and J. E. Mooij, Selective darkening of degenerate transitions demonstrated with two superconducting quantum bits, Nature Phys.~{\bf 6}, 763 (2010).

\bibitem{Moore} C. Moore, Quantum circuits: fanout, parity, and counting, arXiv:quant-ph/9903046.

\bibitem{Nielsen} M. A. Nielsen and I. L. Chuang, {\it Quantum Computation and Quantum Information} (Cambridge University Press, New York, 2000).

\bibitem{Selinger} P. Selinger, Quantum circuits of T-depth one, Phys. Rev. A {\bf 87}, 042302 (2013).

\bibitem{JWfootnote} We note that this approach for compressing the information about a control operator from being a product of many single-qubit operators to a single control qubit operator can be used in the quantum simulation of fermionic systems in multiple dimensions using the Jordan-Wigner transformation \cite{Barends}.

\bibitem{Inada} T. Inada, W. Jang, Y. Iiyama, K. Terashi, R. Sawada, J. Tanaka, S. Asai, Measurement-free ultrafast quantum error correction by using multi-controlled gates in higher-dimensional state space, arXiv:2109.00086.

\bibitem{GroverSearch} L. K. Grover, A fast quantum mechanical algorithm for database search, Proceedings of the 28th Annual ACM Symposium on the Theory of Computing, (1996).

\bibitem{Barends} R. Barends {\it et al.}, Digital quantum simulation of fermionic models with a superconducting circuit, Nature Commun. {\bf 6}, 7654 (2015).

\end{thebibliography}
\end{document}